\documentstyle[preprint,aps,epsfig]{revtex}
\def\plb#1#2#3#4{#1, Phys. Lett. {\bf #2B}, #3 (#4)}
\def\npb#1#2#3#4{#1, Nucl. Phys. {\bf B#2}, #3 (#4)}
\def\prd#1#2#3#4{#1, Phys. Rev. {\bf D#2}, #3 (#4)}
\def\prl#1#2#3#4{#1, Phys. Rev. Lett. {\bf #2}, #3 (#4)}

\def\rep#1#2#3#4{#1, Phys. Rep. {\bf #2}, #3 (#4)}

\begin{document}

\draft
\preprint{
\begin{tabular}{r}
KAIST-TH 2000/13
\\
hep-ph/0011275 
\\ \\ \\
\end{tabular}
}

%%%%%%%%%%%%%%%%%%%%%%%%%%%%%%%%%%%%%%%%%%%%%%%%%%%%%%%%%%%%%%%%%%%%%%%%%%%%%%%

\title{Bounds on the Mass and Coupling Constant of 
Radion in the Randall-Sundrum Theory}

\author{
Saebyok Bae
\thanks{E-mail: dawn@muon.kaist.ac.kr}
and Hong Seok Lee
\thanks{E-mail: hslee@muon.kaist.ac.kr}
}

\address{
Department of Physics, Korea Advanced Institute of Science and
Technology \\
Taejon 305-701, Korea \\
}

\maketitle

\tighten

\begin{abstract}
Assuming the Goldberger-Wise mechanism, we investigated the effective
potential at the one-loop level in the Randall-Sundrum theory.
We found the lower and upper bounds of the radion mass $m_{\phi}$
and the radion-SM coupling constant, 
$0.8~{\rm GeV} \lesssim m_{\phi} \lesssim 260~{\rm GeV}$ and
$1400~{\rm GeV} \lesssim \Lambda_{\phi} \lesssim 1500~{\rm GeV}$ for the
one-loop level potential. These bounds were determined 
from two constraints of ${\rm warp~ factor}=O(M_W/M_{\rm Pl})$ and 
${\rm Higgs}~ vev \simeq 246$ GeV, which can produce strong  
bounds of $m_\phi$ and $\Lambda_{\phi}$.
It is phenomenologically important that the
one-loop allowed upper bound of the radion mass is about five times
larger than the tree-level one, but the radion is still lighter
than the Kaluza-Klein modes.
\end{abstract} 

%\tighten 

\newpage 
%%%%%%%%%%%%%%%%%%%%%% 
\section{Introduction} 
%%%%%%%%%%%%%%%%%%%%%% 
Recently it was known that the large difference between the Planck scale
$\Lambda_{\rm Pl} \sim 10^{18}$ GeV and the electroweak scale 
$\Lambda_{\rm EW}\sim 10^2$ GeV (so-called the
hierarchy problem) can be produced by introducing the compact 
extra dimension(s). There are two models with extra dimension(s). One is the 
ADD model (large extra-dimension
model) \cite{ADDA}, where the extra dimensions are below
$O(1) \, mm$ ($\gg 1/M_W$) and the extra submanifold is 
at least two dimensional. 
Although the ADD model resolves the hierarchy problem, it causes another 
hierarchy problem between the electroweak scale and the radius of the extra 
dimensions. This comes from the power law behavior of the hierarchy, that
is, $\Lambda_{\rm EW}= (R \Lambda_{\rm Pl})^{-n/(2+n)} \Lambda_{\rm Pl}$, 
where $R$ and $n$ are the typical radius and the dimension of the extra
submanifold, respectively. 
The other is the Randall-Sundrum (RS) model (small extra-dimension model)
\cite{RS}, where the classical RS metric is $ds_{\rm
RS}^2=e^{-2kr_c|y|}\eta_{\mu \nu}dx^{\mu}dx^{\nu}-r_c^2dy^2$ and the extra
manifold is a one-dimensional orbifold $S^1/Z_2$. $k$ is smaller than 
the 5-dimensional Planck mass $M_{(5)}\sim M_{\rm Pl}
\simeq 2\times10^{18}$ GeV, $\pi r_c$ is 
the interbrane distance, and the hidden and visible branes are at $y=0$ and 
$y=\pi$, respectively. The electroweak scale can be 
derived from the Planck scale via $\Lambda_{\rm EW}=e^{-k r_c \pi} 
\Lambda_{\rm Pl}$, where $e^{-k r_c \pi}$ is the geometrical warp
factor appearing in the metric
of our brane. The unwanted extra hierarchy problem does not exist due to 
the exponential dependence of the hierarchy on the radius $r_c$ unlike the 
ADD theory.  

When a kind of modulus stabilization with a bulk scalar is added to the RS 
theory (since it is impossible to determine the value of the interbrane 
distance~$\sim O(10/M_{(5)})$ from the dynamics of the ``original" RS model)
and a kind of dimensional reduction is done, the modulus field (radion) 
couples to the Standard Model (SM) fields like 
$\bar{\phi} T^{\mu}_{\mu}/\Lambda_{\phi}$
\cite{{GW11457},{Csaki}}, where $\bar{\phi}$ is the physical field of the
radion, $\Lambda_{\phi}$ is a dimensionful coupling constant of order
$\Lambda_{\rm EW}$, and $T^{\mu}_{\mu}$ is the trace of the
energy-momentum tensor of the SM fields. The magnitudes of $\Lambda_{\phi}$ 
and the radion mass $m_{\phi}$
are phenomenologically important, since they are related to decay rates and
cross sections. There are two kinds of considerations on these parameters until
now. One is experimental, and the other is theoretical. The experimental
consideration is related to the experimentally measured quantities, for
example, the decay rates and  production cross sections of radion, and the 
radion loop corrections to the weak mixing angle 
\cite{{Giudice},{Mahanta},{Bae},{Kim},{Mahanta6006}}. It is noteworthy that 
the branching ratio of the radion into two gluons is the largest for 
$m_\phi \lesssim 160$ GeV, although it comes from the trace anomaly at the 
one-loop level. The recent L3 results on the SM Higgs search excluded  
$\Lambda_\phi \lesssim 600$ GeV for $m_\phi \lesssim 100$ GeV \cite{Bae}. 
The constraint from the weak mixing angle with errors of order 
$O(0.1)\%$ can exclude $\Lambda_{\phi} \sim 500$ GeV for $m_{\phi} 
\sim 500$ GeV \cite{Kim}. And the lower bound of the radion mass is about 
$O(1)$ GeV from limits of neutrino phenomenology and TASSO and CLEO on 
inclusive B decays \cite{Mahanta6006}. The theoretical one is on the 
perturbative unitarity \cite{{Bae},{Mahanta4128}}. The perturbative 
unitarity for $hh \rightarrow hh$ ($h$: SM Higgs boson) is broken for 
relatively small $\Lambda_\phi \lesssim 200$ GeV 
for $m_\phi \sim 600$ GeV \cite{Bae}, and the unitarity bound in the 
$W^+_{L}W^-_{L} \rightarrow h \phi$ process is $m_{\phi} \lesssim 2500$ GeV 
at $\Lambda_\phi=1$ TeV \cite{Mahanta4128}.

In this paper, we have constrained the two parameters up to one-loop level in
terms of two conditions, warp factor $e^{-kr_c\pi}=
O(\Lambda_{\rm EW}/\Lambda_{\rm Pl})$ and ${\rm Higgs}~ vev \simeq 246$ GeV
($vev= {\rm vacuum~ expectation~ value}$). 
%The vacuum stability is related to the fact that the modulus stabilization
%and the electroweak symmetry breaking have to be maintained even though
%the quantum corrections are added to the tree level Lagrangian. 
The constraint, ${\rm warp~ factor}=O(\Lambda_{\rm EW}/\Lambda_{\rm Pl})$, 
was imposed to solve 
the hierarchy problem \cite{RS}. From the above conditions, the one-loop 
allowed bounds of $m_\phi$ and $\Lambda_\phi$
are $0.8~{\rm GeV} \lesssim m_{\phi}\lesssim 260~{\rm GeV}$ and 
$1400~{\rm GeV} \lesssim \Lambda_{\phi} \lesssim 1500~{\rm GeV}$, which can be
stronger than the previous constraints except the lower bound of 
radion mass. It is noteworthy that the one-loop
upper bound of the radion mass $m_\phi$ is about five times larger
than the tree-level one of about 49 GeV, and still smaller than the masses of
the low-lying Kaluza-Klein (KK) modes.

This paper is organized as follows: In Section II, we discuss 
what fields give important contributions to the effective potential,
and analyze the contributions of the scalar and KK mode
sectors and those of the fermion and gauge sectors. In Section III, we
find the bounds on $m_\phi$ and $\Lambda_\phi$ from  
${\rm warp~ factor}=O(\Lambda_{\rm EW}/\Lambda_{\rm Pl})$
and ${\rm Higgs}~ vev \simeq 246$ GeV. The conclusions are in Section IV.

%%%%%%%%%%%%%%%%%%%%%%%%%%%%%%%%%%%%%%
\section{One-Loop effective potential}
%%%%%%%%%%%%%%%%%%%%%%%%%%%%%%%%%%%%%%
When the hierarchy problem is solved with only the one-loop corrections
of bulk fields, they generate a very light radion \cite{Garriga}, which is
excluded by experiments of neutrino phenomenology 
and $B$ decays \cite{Mahanta6006}. Therefore, the 
stabilization mechanism at the tree level is required \cite{{Garriga},{GR}}.
The mechanism proposed by Goldberger and Wise \cite{GW7447} can be a promising
one, because they stabilized the modulus without any severe fine-tuning of the
parameters in the full theory. In the Goldberger-Wise stabilization mechanism, 
there is a bulk scalar field $\Phi(x^\mu, y)$ which has large quartic 
self-interactions on the hidden and visible branes alone, and a 
$y$-dependent $vev$ ${\tilde \Phi}(y)$. After replacing the field 
$\Phi(x^\mu, y)$ in the original Lagrangian with its $vev$ ${\tilde \Phi}(y)$ 
and integrating the Lagrangian over $y$, we have the modulus stabilizing 
potential. The typical energy of this mode is the expectation
value of an operator $p_y=-\frac{1}{r_c}\frac{\partial}{\partial y}$ to 
${\tilde \Phi}(y)$ of order $O(10^{-2})k$, 
%$1/({\rm interbrane ~distance})=1/\pi r_c=O(10^{-2}M_{(5)})$, 
which makes the effective 
gravitational coupling constant $({\rm Energy}/M_{(5)})$ of the bulk scalar 
about $O(10^{-2})$ by the 5-dimensional Planck suppression. Thus the quantum 
correction of the 5-dimensional graviton to the bulk scalar $vev$ 
${\tilde \Phi}(y)$ is unimportant in the full theory. And the quantum 
corrections from the brane self-interactions of $\Phi$ can shift the $vev$s of
the bulk scalar at the branes, $v_v$ and $v_h$, where 
$v_v={\tilde \Phi}(\pi)$ and $v_h={\tilde \Phi}(0)$. These corrections can 
change only the coefficients of the radion potential and be equivalent to
rescaling of the coefficients.

The 4-dimensional graviton coupling to the radion is ${\cal L}_{\rm grav}=
-(2M_{(5)}^3/k) e^{-2 \phi/\Lambda_{\phi}}{\cal R}_{(4)}$ 
\cite{{RS},{GW11457}}, where ${\cal R}_{(4)}$ 
is the 4-dimensional Ricci scalar. When the radion $\phi$ is replaced with 
$\phi=\phi_0+\bar{\phi}$, where $\phi_0$ is the $vev$ of $\phi$,
%making $(2M_{(5)}^3/k) e^{-2 \phi_0/\Lambda_{\phi}} %={\Lambda_{EW}}^2$, 
the one-graviton coupling constant to the physical field of the radion 
$\bar{\phi}$ is Planck-suppressed, $O(\Lambda_{\rm EW}/M_{\rm Pl}) \ll 1$,
since ${\cal R}_{(4)}$ has linear terms 
of the graviton field $h_{\mu \nu}(x)=(g_{\mu \nu}(x)-\eta_{\mu
\nu})/\sqrt{8 \pi G_N}$ in the weak field approximation. So we can neglect 
the quantum effects of the 4-dimensional graviton. There are low-lying
KK modes of the bulk scalar and the bulk graviton with masses 
of $O(1)$ TeV and non-renormalizable couplings suppressed by TeV scale 
\cite{GW7218}. The effects of these low-lying KK modes on the electroweak 
scale physics are absorbed into the renormalizations of the Planck and TeV 
brane tension terms $V_{\rm Planck}
+V_{\rm TeV} e^{-4 \phi/\Lambda_\phi}$ \cite{GR}. Since the radion has
highly non-renormalizable self-interactions such as $V_\phi$ in 
Eq.~(\ref{Vphi}), we treated the radion self-interactions 
only at the tree level. For example, the
non-renormalizable quantum gravity has a serious problem that the
one-loop effective action on the mass-shell is dependent on the gauge fixing
parameters \cite{ichinose}, breaking the DeWitt-Kallosh theorem \cite{dewitt}.
Therefore, we consider the loop corrections of the KK modes of bulk 
fields and the SM particles to the radion effective potential in our brane. 

%%%%%%%%%%%%%%%%%%%%%%%%%%%
\subsection{Scalar and KK mode sectors}
%%%%%%%%%%%%%%%%%%%%%%%%%%%
To make its kinetic term canonically normalized, we rescale the Higgs field 
$H(x) \rightarrow e^{kb(x)/2}H(x)$ \cite{Csaki}, where 
$kb(x)/2=\phi(x)/\Lambda_\phi$.
After this rescaling, the scalar sector has the Lagrangian density
\cite{{GW11457},{Csaki}} 
\begin{eqnarray}
\label{Lscalar}
{\cal L}_{\rm scalar} &=& D_{\mu}H^{\dagger}D^{\mu}H 
+\frac{1}{2}(\partial \phi)^2 - V_H(H, \phi) - V_{\phi}(\phi)
\nonumber 
\\
&&+
\frac{\partial_{\mu}\phi}{\Lambda_{\phi}}(H^{\dagger}D^{\mu}H + h.c.)
+\left(\frac{\partial\phi}{\Lambda_{\phi}}\right)^2 H^{\dagger}H .
\end{eqnarray}
The potentials $V_H$ \cite{Csaki} and $V_{\phi}$ \cite{GW7447} are
\begin{equation}
\label{Vh}
V_H(H, \phi) = \lambda \left( H^{\dagger}H-
\frac{1}{2}v_0^2e^{- 2\phi/\Lambda_{\phi}} \right)^2 ,
\end{equation}
and
\begin{eqnarray}
\label{Vphi}
V_{\phi}(\phi) &=& 4ke^{-4 \phi/ \Lambda_{\phi} } \left(v_v-v_h
e^{-\epsilon \phi/ \Lambda_{\phi}} \right)^2
\left(1+\frac{\epsilon}{4} \right) 
\nonumber
\\
&& - k \epsilon v_h 
e^{-(4+\epsilon) \phi/ \Lambda_{\phi}} \left(2 v_v - v_h
e^{-\epsilon \phi/ \Lambda_{\phi}} \right) , 
\end{eqnarray}
where $\epsilon \simeq m^2/4k^2$ is about $1/40$ ($k/m$ is of order unity) 
%in the Goldberger-Wise limit of large quartic coupling constants on the branes
to solve the hierarchy problem \cite{GW7447}. The ranges of $v_0$, $v_v^{2/3}$, 
$v_h^{2/3}$, $k$, and the bulk scalar mass $m$ are of order $O(0.1)M_{(5)}$ 
in order to avoid a large hierarchy in the RS theory and 
to maintain the classical RS metric or, equivalently, 
neglect the loop corrections of the 5-dimensional 
quantum gravity. In the Landau gauge, three massless would-be Goldstone
bosons appeared in the calculation unlike the unitary gauge. 
The fifth term in Eq.~(\ref{Lscalar}) is the mixing term of radion 
and Higgs fields with two derivatives. It has to be considered in the 
canonical normalizations of the kinetic terms of the radion and Higgs 
fields \cite{Csaki}. The effective potential from the scalar sector 
in the $\overline{\rm MS}$ scheme is
\begin{eqnarray}
\label{Vscalar}
V_{\rm eff}^{\rm scalar}(h, \phi) &=&
V_{\rm tree}(h, \phi) + V_{\rm 1 \, loop}^{\rm scalar}(h, \phi)
\nonumber
\\
&=&
V_{\phi}(\phi)+\frac{\lambda}{4} \left(h^2 - v_0^2 
e^{- 2 \phi/\Lambda_{\phi}} \right)^2
\nonumber
\\
&&
\,+\frac{1}{4(4\pi)^2} 
\left\{\lambda^2 \left(3h^2 - v_0^2 e^{- 2 \phi/\Lambda_{\phi}} \right)^2
\left( \log \left[ \lambda \left(3h^2 - v_0^2 e^{- 2 \phi/\Lambda_{\phi}} 
\right)/{\mu}^2 \right] - \frac{3}{2} \right) 
\right.
\\
&&
\left.
~~~~~~~~~~~~~+ 3 \lambda^2 \left(h^2 - v_0^2 e^{- 2 \phi/\Lambda_{\phi}} \right)^2
\left( \log \left[ \lambda \left(h^2 - v_0^2 e^{- 2 \phi/\Lambda_{\phi}} 
\right)/{\mu}^2 \right] - \frac{3}{2} \right) \right\} , 
\nonumber
\end{eqnarray}
where $\mu$ is a renormalization scale.   
The term with a coefficient $\lambda^2$ in Eq.~(\ref{Vscalar}) came from 
the physical Higgs field, and the 3 Goldstone bosons $G^0$, $G^+$ and $G^-$ 
gave the next term with a coefficient $3\lambda^2$. The KK modes originated 
from the bulk scalar can couple to the
Higgs field with trilinear couplings \cite{CsakiG} which give the effective
quartic term of the Higgs field $\frac{c_{\rm KK}}{4}h^4$ at the tree level.
The contributions of the low-lying KK modes to $c_{\rm KK}$ are much smaller 
than the SM one-loop corrections in Eq.~(\ref{Vscalar}), and thus can be 
neglected.

But the one-loop corrections from the KK modes of the bulk scalar
and graviton can contribute considerably to the radion potential as 
$V_{\rm 1 \, loop}^{\rm KK}(\phi)=\delta V_{\rm Planck}^{\rm KK}
+ \delta V_{\rm TeV}^{\rm KK} 
e^{-4\phi/\Lambda_{\phi}}$ \cite{{Garriga},{GR}}, where 
$\delta V_{\rm Planck}^{\rm KK}$ and $\delta V_{\rm TeV}^{\rm KK}$ are
shifts from the
classical Planck and TeV brane tensions $V_{\rm Planck}=-V_{\rm TeV}=
24 M_{(5)}^3 k >0$ \cite{RS} respectively, belonging to the tension
shifts $\delta V_{\rm Planck}$ and $\delta V_{\rm TeV}$ of $V_{\Lambda}(r_c)
= \delta V_{\rm Planck} + \delta V_{\rm TeV} 
e^{-4kr_c \pi}$ in Ref.~\cite{GW7447}. For small $\delta V_{\rm TeV}$,
the addition of $V_\Lambda$ to the potential $V_\phi$ in Eq.~(\ref{Vphi})
gives a minimum for large $k r_c$, and the adjustment of the tension shift
$\delta V_{\rm Planck}$ on the Planck brane can make the effective
four-dimensional cosmological constant vanish \cite{{GW7447},{Garriga}}.
Since the shift of the TeV brane tension $\delta V_{\rm TeV}^{\rm KK}$ 
is absorbed into the renormalization of the TeV brane tension term \cite{GR}
$S_{\rm tension}^{\rm TeV} = -\int d^4 x \sqrt{-g_{\rm TeV}}\,V_{\rm TeV}$ 
\cite{{RS},{GW7447}},
it can not be obtained from calculation and can only be determined by 
a renormalization condition relating it to observable
quantities \cite{Garriga}. Because the 4-dimensional cosmological
constant of order $10^{-120} M_{\rm Pl}^4 \ll \Lambda_{\rm EW}^4$ 
\cite{Riess} is regarded as the minimum value of the effective potential 
$V_{\rm eff}(h,\phi)$ of the RS theory, the value of the potential 
can be approximated as zero  at the observed values of $h$ and $\phi$,
which should be a stable minimum point of the potential \cite{Garriga}.
The two renormalization conditions are expressed as  
\begin{equation}
\label{rencond}
V_{\rm eff}=0
~~~{\rm and} ~~~
\frac{\partial V_{\rm eff}} {\partial h}=
\frac{\partial V_{\rm eff}} {\partial \phi}= 0 
% ~~~{\rm at~ the ~ observed ~ values~ of~ {\it h} ~ and~ \phi}.
\end{equation}
at the observed values of $h$ and $\phi$.
From these first and second conditions, we can determine the shifts 
$\delta V_{\rm Planck}$ and $\delta V_{\rm TeV}^{\rm KK}$ respectively, if
all the other parameters are known. The size of the shift 
$\delta V_{\rm TeV}^{\rm KK}$ can be of order $O(\Lambda_{\rm Pl}^4/100)$
much smaller than $|V_{\rm TeV}|$.
%due to the second condition. since the tree-level tensions are exactly 
%canceled by the bulk cosmological constant. 
It is noteworthy that the sign of the KK mode contribution 
is important in the vacuum 
stability since it can change the shape of potential. If 
$\delta V_{\rm TeV}^{\rm KK}$ is fixed to be negative and 
$|\delta V_{\rm TeV}^{\rm KK}|$ increases, the potential becomes deeper 
and has more parameter points which give stable vacua. But when the sign 
is reversed, the number of the allowed points gets smaller, and thus the
negative $\delta V_{\rm TeV}^{\rm KK}$ can be favored in terms of stability.

%%%%%%%%%%%%%%%%%%%%%%%%%%%%%%%%%%%%%%%%%%
\subsection{Fermion and Vector sectors}
%%%%%%%%%%%%%%%%%%%%%%%%%%%%%%%%%%%%%%%%%%
Since we used the $x^\mu$-dependent rescaling $\psi(x) \rightarrow e^{3kb(x)/2}
\psi(x)$ for a fermion $\psi$, the radion-fermion interaction Lagrangian 
is ${\cal{L}}_{\phi \psi \bar{\psi}}=i(3/2\Lambda_\phi) \bar{\psi} \gamma^\mu
\psi \partial_\mu \phi$ \cite{Csaki}. The Lagrangians for the gauge bosons
are similar to that of the Higgs boson. The background field method 
\cite{PeSc} and the tadpole method 
\cite{Sher} showed easily that the fermions and gauge bosons of the Standard
Model do not give any one-loop contributions to the radion potential although
the interaction Lagrangians are non-renormalizable.
%But these corrections can be interpreted as the loop corrections
%to the Higgs potential, since the masses are proportional to the
%electroweak-scale Higgs $vev$, $v_{\rm Higgs} e^{-kb_0/2}$ 
%($b_0 \equiv {\rm $vev$~of}~b(x)=2\pi r_c$), where $v_{\rm Higgs}$ is the 
%Higgs $vev$ of order $M_{Pl}$ in the RS theory. (Since $v_{\rm Higgs}
%e^{-kb_0/2}$ corresponds to the field operator $h_0(x)e^{-kb(x)/2}$, where
%$h_0(x)$ is the neutral Higgs field before the rescaling, the brane tension
%term $V_{TeV}e^{-2kb(x)}$ can not be the counter term of one-loop corrections
%of this operator.) Therefore, only the Higgs 
%potential received loop corrections from those particles effectively. 
Therefore, only the Higgs potential received the loop corrections from these
particles. The contributions of the fermion and vector sectors to the 
effective potential are 
\begin{eqnarray}
\label{Vfervec}
V_{\rm eff}^{\rm fer+vec}(h) &=&
V_{\rm 1 \, loop}^{\rm fermion}(h) + V_{\rm 1 \, loop}^{\rm vector}(h)
\nonumber
\\
&=&
\frac{1}{(4 \pi)^2} \left\{
-3T^2\left(\log\frac{T}{\mu^2}-\frac{3}{2} \right)
+\frac{3}{2}W^2 \left(\log\frac{W}{\mu^2}-\frac{5}{6} \right)
+\frac{3}{4} Z^2 \left(\log\frac{Z}{\mu^2}-\frac{5}{6} \right)
\right\} ,
\end{eqnarray}
where $T = \frac{1}{2}Y_t^2h^2$, $W = \frac{1}{4}g^2h^2$, 
and $Z = \frac{1}{4}(g^2+{g'}^2)h^2$ \cite{JoSt}. The $g$ and $g'$ are the 
gauge coupling constants of ${\rm SU(2)}_L \times {\rm U(1)}_Y$, and the 
$Y_t$ is the top quark Yukawa coupling 
constant. Since the fermion contribution is proportional to $(\rm Yukawa~
coupling~ constant)^4$, only the top quark contribution is considered.  
The difference of constants in the parentheses of the fermion sector and
the gauge sector came from the gauge
dependent term $(1-\xi)k^{\mu}k^{\nu}/(k^2-\xi m^2)$ ($\xi=0$) of 
the gauge boson propagator. This can be seen qualitatively and easily 
in the tadpole method \cite{Sher}. Therefore, adding the contribution of the
KK modes of the bulk fields, the final effective potential up to the one-loop 
level is  
\begin{eqnarray}
\label{Veff}
V_{\rm eff}(h, \phi)
&=&
V_{\rm tree}(h, \phi) + V_{\rm 1 \, loop}^{\rm scalar}(h, \phi)
+V_{\rm 1 \, loop}^{\rm fermion}(h) + V_{\rm 1 \, loop}^{\rm vector}(h) 
+V_{\rm 1 \, loop}^{\rm KK}(\phi)
\nonumber
\\
&=&
V_{\phi}(\phi)+\frac{\lambda}{4} \left(h^2 - v_0^2 
e^{- 2 \phi/\Lambda_{\phi}} \right)^2
\nonumber
\\
&&
\,+\frac{1}{4(4\pi)^2} 
\left\{ 
\lambda^2 \left(3h^2 - v_0^2 e^{- 2 \phi/\Lambda_{\phi}} \right)^2
\left( \log \left[ \lambda \left(3h^2 - v_0^2 e^{- 2 \phi/\Lambda_{\phi}} 
\right)/{\mu}^2 \right] - \frac{3}{2} \right) \right\}  
\nonumber
\\
&&
\left.
~~~~~~~~~~~~~+ 3 \lambda^2 \left(h^2 - v_0^2 e^{- 2 \phi/\Lambda_{\phi}} 
\right)^2 \left( \log \left[ \lambda \left(h^2 - v_0^2 e^{- 2 \phi/
\Lambda_{\phi}} \right)/{\mu}^2 \right] - \frac{3}{2} \right) \right\}  
\\
&&
\, +\frac{1}{(4 \pi)^2} \left\{
-3T^2\left(\log\frac{T}{\mu^2}-\frac{3}{2} \right)
+\frac{3}{2}W^2 \left(\log\frac{W}{\mu^2}-\frac{5}{6} \right)
+\frac{3}{4} Z^2 \left(\log\frac{Z}{\mu^2}-\frac{5}{6} \right)
\right\}
\nonumber
\\
&&
\,+ \delta V_{\rm Planck}^{\rm KK} + \delta V_{\rm TeV}^{\rm KK}
e^{-4\phi/\Lambda_{\phi}} .
\nonumber
\end{eqnarray}
Note that all the one-loop
corrections are of the same order $O(\Lambda_{\rm EW}^4/100)$ for our parameter
range.  The $vev$s $v$ and $\phi_0$ of the Higgs neutral component field 
$h(x)$ and the radion field $\phi(x)$ are determined by the stationary
condition of the effective 
potential, when the fields $h$ and $\phi$ are assumed to be independent of 
spacetime coordinates. It is reasonable that we should consider the region 
of $v$ and $\phi_0$ where the validity of perturbation and the condition of 
real $V_{\rm eff}$ are satisfied.

%%%%%%%%%%%%%%%%%%%%%%%%%%%%%%%%%%%%%%%%%%%%%%%%%%%%%%%%%%%%%%%%%%%%%%%%%%%%
\section{Bounds on $m_\phi$ and $\Lambda_{\phi}$}
%%%%%%%%%%%%%%%%%%%%%%%%%%%%%%%%%%%%%%%%%%%%%%%%%%%%%%%%%%%%%%%%%%%%%%%%%%%%
%Although a scalar field ({\it e.g.} SM Higgs boson) has a non-vanishing $vev$ 
%at the tree level, it depends on the parameters of the system that the $vev$ 
%of the scalar is still non-zero at the loop levels. If the loop corrections 
%at some parameters make the $vev$ vanish, or destroy the tree level $vev$, 
%then the symmetry of the system ({\it e.g.} electroweak symmetry) is not 
%spontaneously broken. 
In the Randall-Sundrum theory, the extra dimension must have a finite size
about $O(10/M_{\rm Pl})$ to solve the hierarchy problem \cite{{RS},{GW7447}}, 
which gives the constraint, warp factor $e^{-kr_c\pi}=O(\Lambda_{\rm
EW}/\Lambda_{\rm Pl})$. 
%Therefore, the modulus has to be stabilized at a finite nonzero $vev$. This
%is a constraint for the RS standard model. 
And in the Standard Model, the $vev$ of the Higgs field is about 246 GeV. This
can give a constraint on the Higgs $vev$ of the effective potential in the
RS model. But when the quantum corrections of radion are 
added to the SM particles, the experimentally 
determined $vev$ of the Higgs field may be much different from the SM value, 
that is, the corrections may destroy the above constraint.    
When we consider the quantum corrections of the radion to the masses of the SM
particles using the Lagrangian ${\cal{L}}_{\phi \psi \bar{\psi}}=
i(3/2\Lambda_\phi) \bar{\psi} \gamma^\mu \psi \partial_\mu \phi$, 
the pole masses of leptons and heavy quarks, for example, are 
\begin{eqnarray}
\label{m_f}
m_f^{\rm pole} 
&\simeq&
m_f^{\rm SM} 
\left\{ 
1+\frac{1}{(4 \pi)^2} 
\left(\frac{m_f}{\Lambda_{\phi}}\right)^2  
\int_0^1 dx 
\left[
\frac{9}{4}(3x-1)\frac{\Delta(m_f^{\rm pole})}{m_f^2}
\right.
\right.
\nonumber
\\
&&
\left.
\left.
+\left(1+x-\frac{9}{4}(1+x)(1-x)^2 \left(\frac{m_f^{\rm pole}}{m_f} \right)^2
-\frac{9}{2}(3x-1)\frac{\Delta(m_f^{\rm pole})}{m_f^2}
\right)
\ln \frac{\Delta(m_f^{\rm pole})}{\mu^2}
\right]
\right\} , 
\end{eqnarray}
where $m_f^{\rm SM}$ are the renormalized masses with the SM loop corrections
alone, $m_f$ are the mass parameters in the renormalized Lagrangian, 
$\mu$ is a renormalization scale, and $\Delta(m_f^{\rm pole})= 
-x(1-x)(m_f^{\rm pole})^2+xm_{\phi}^2 +(1-x)m_f^2$. 
If $\Lambda_{\phi} \lesssim O(m_f/4 \pi)$, then the radion correction can be 
large so that the shift of the Higgs $vev$ from the SM one can be large. (This
argument is similarly applied to the gauge bosons and the Higgs boson.)
So the condition $v\simeq 246$ GeV can be spoilt for  
$\Lambda_{\phi} \lesssim O(m_f/4 \pi)$. But this small range is excluded  
by the constraint from the weak mixing angle that 
$\Lambda_\phi \lesssim 300$ GeV are excluded for
$m_\phi \le 1000$ GeV \cite{Kim}. Therefore, it can be natural that 
the physical pole mass ($e.g.$ $m_f^{\rm pole}$) is almost equal to the 
SM mass ($e.g.$ $m_f^{\rm SM}$). This means that the 
Higgs $vev$ $\simeq 246$ GeV can be a physical constraint on our 
effective potential for the natural range of $\Lambda_{\phi}$.  
The radion corrections are typically smaller than $O(0.1)\,\%$ of 
$m_f^{\rm SM}$, so we can expect that the errors of the Higgs $vev$ from the
radion contributions are smaller than $O(0.1)\,\%$. 

The procedure to determine $m_\phi$ and $\Lambda_\phi$ is as follows.
First, we find a parameter point of ($v_0,~ v_v,~ v_h,~ m,~
\delta V_{\rm TeV}^{\rm KK}$) which gives a stable vacuum satisfying the 
two constraints, warp factor $e^{-kr_c\pi}=O(\Lambda_{\rm EW}
/\Lambda_{\rm Pl})$ and Higgs $vev$ $v \simeq 246$ GeV. Note that  
$\delta V_{\rm TeV}^{\rm KK}$ is not considered for the tree-level potential.
Of course, the point has to be in the region where perturbation is 
valid and $V_{\rm eff}$ is real. Next, the values of $m_\phi$ and
$\Lambda_\phi$ are determined from that point by means of 
equations $m_{\phi}^2=\frac{\partial^2 V_\phi} {\partial \phi^2}(\phi_0)$
and $\Lambda_{\phi}=\sqrt{6}M_{\rm Pl}e^{-kb_0/2}$. Since when a minimum
point is found, $\delta V_{\rm Planck}^{\rm KK}$ can be determined trivially 
via the first equation in Eq.~(\ref{rencond}), we do not include
the parameter $\delta V_{\rm Planck}^{\rm KK}$.
Because the RS metric $ds^2_{\rm RS}$ is a solution of the ``classical"
5-dimensional Einstein's equation, in order to maintain the classical metric
we choose  parameter ranges where the loop effects of the quantum gravity
can be neglected.

\begin{figure}
\centerline{\epsfxsize=9cm \epsfbox{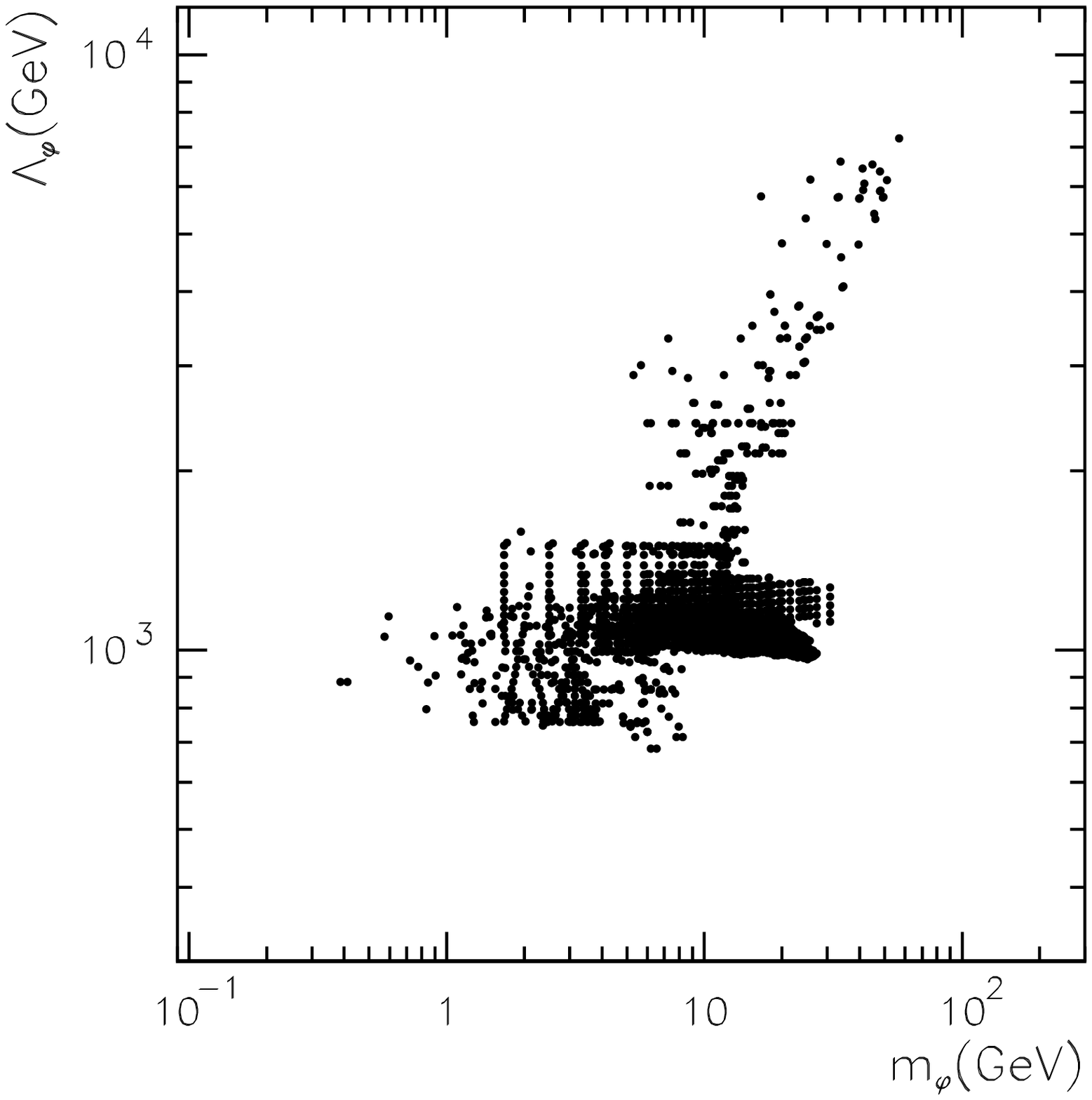}}  
\caption{The allowed parameter points in $m_\phi$ and $\Lambda_\phi$
space for the tree level potential ($M_{(5)}=0.8M_{\rm Pl}$ and 
$m_h=125$ GeV).}
\label{tree1}
\end{figure}

\begin{figure}
\centerline{\epsfxsize=9cm \epsfbox{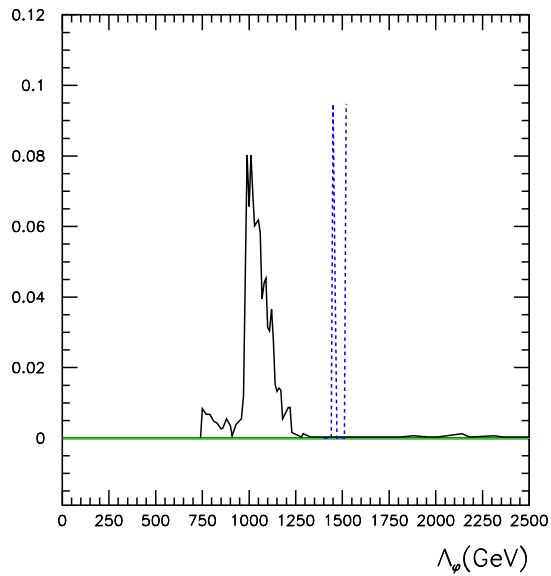}} 
\caption{The $y$ coordinates of the graphs are proportional to 
the numbers of allowed parameter points. The solid and dotted
lines denote the allowed points for the tree and one-loop level
potentials, respectively. The tree-level points with 
$\Lambda_\phi > 2500$ GeV are not shown.} 
%($M_{(5)}=0.8M_{\rm Pl}$ and $m_h=125$ GeV).}
\label{lam-num}
\end{figure}

For the tree level potential, the allowed region for the 5-dimensional 
Planck mass of a typical size $M_{(5)}=0.8M_{\rm Pl}$ and the Higgs
mass $m_h=125$ GeV is $0.4~ {\rm GeV} \lesssim m_{\phi} \lesssim 49~ {\rm
GeV}$ and $760~ {\rm GeV}\lesssim \Lambda_{\phi} \lesssim 
5900~ {\rm GeV}$ 
from Fig.~\ref{tree1}, where $\Lambda_{\phi}=\sqrt{6}M_{\rm Pl}e^{-kb_0/2}$
and $m_{\phi}^2= \frac{\partial^2 V_\phi}{\partial \phi^2}(\phi_0)$. 
Because of the exponential dependence of 
$\Lambda_{\phi}= \sqrt{6} M_{\rm Pl}e^{-kb_0/2}$, the large change of 
$\Lambda_{\phi}\sim O(100)\,\%$ corresponds to the very small change of 
$kb_0/2\sim O(1)\,\%$, whose central value is about 36 ($kr_c \simeq 12$).
This means that the range of the radion $vev$ $\phi_0$ is very narrow.
From Fig.~\ref{lam-num}, most of the allowed parameter points 
(about 93 percent) are concentrated around the line $\Lambda_\phi=1070$ GeV,
and thus the point with a smaller or larger $\Lambda_\phi$ is rare. 
Therefore, the {\it naturally} allowed region can be much 
narrower in $\Lambda_\phi$ than the above allowed one.

\begin{figure}
\centerline{\epsfxsize=9cm \epsfbox{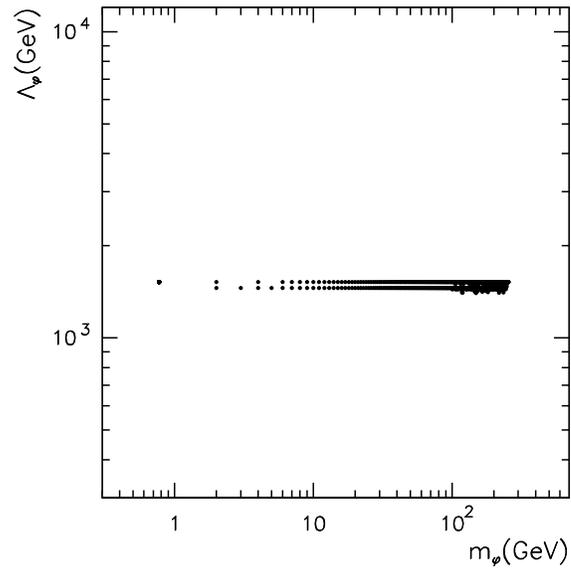}}  
\caption{The allowed points for the one-loop effective potential 
($M_{(5)}=0.8M_{\rm Pl}$ and $m_h=125$ GeV).}
\label{loop1}
\end{figure}

For the one-loop effective potential, the allowed region for the 5-dimensional 
Planck mass $M_{(5)}=0.8M_{\rm Pl}$ and the Higgs mass $m_h=125$ GeV is 
$0.8~{\rm GeV} \lesssim m_{\phi} \lesssim 260~{\rm GeV}$ and 
$1400~{\rm GeV} \lesssim \Lambda_{\phi} \lesssim 1500~{\rm GeV}$ 
from Fig.~\ref{loop1} (the central value of $kb_0/2$
is still about 36). 
%Note that the blank region below the line 
%$\Lambda_{\phi} \approx 6 m_{\phi}$ is forbidden in Fig.~\ref{loop1}. 
From Fig.~\ref{lam-num},  almost all the data are focused at
$\Lambda_\phi=1490$ GeV like the tree level case. Therefore, we can conclude
that the {\it naturally} allowed regions are similar for the tree and one-loop
cases, and there is a small shift of the central value of 
$\Lambda_\phi$ (or equivalently $\phi_0$) due to the one-loop
corrections. When $\delta V_{\rm TeV}^{\rm KK}$ and other
parameters are changed continuously, the allowed region in Fig.~\ref{loop1}
can be broader in $\Lambda_\phi$. From numerical analysis, we have found 
that only the negative values of $\delta V_{\rm TeV}^{\rm KK}$ 
can be allowed. As discussed in Section II, negative values of the 
tension shift $\delta V_{\rm TeV}^{\rm KK}$ produce many parameter
points which were not allowed at the tree level, and considerable parts
of these new points make the radion mass sufficiently larger
than the tree-level upper bound of the mass. 
It is phenomenologically noteworthy that the one-loop upper bound of the
radion mass $m_\phi$ is rather larger than the tree-level one by
about five times.
But the radion is still the first signal of the RS theory lighter than the
lowest-lying KK mode with a mass of order $O(1)k e^{-kb_0/2} 
\simeq 0.8 \Lambda_\phi$ \cite{{GW7218},{GhPo}}, because the radion mass is
smaller than about $260$ GeV. And the branching ratios of the radion into
gluon or $W$ boson pairs are dominant according to the mass $m_\phi$ 
\cite{{Giudice},{Bae}}.

The tree level perturbative unitarity can not be stronger than the
two constraints, warp factor $e^{-kr_c\pi}=O(\Lambda_{\rm EW}
/\Lambda_{\rm Pl})$ and Higgs $vev$ $v \simeq 246$ GeV, in constraining
$m_{\phi}$ and $\Lambda_{\phi}$, since the 
unitarity bound in the $hh \rightarrow hh$ process is $\Lambda_{\phi} 
\gtrsim 300~{\rm GeV}$ at $m_{\phi} \lesssim 1000 ~{\rm GeV}$ \cite{Bae} and
the bound in the $W^+_{L}W^-_{L} \rightarrow h \phi$ process is $m_{\phi}
\lesssim 2500$ GeV at $\Lambda_\phi=1$ TeV \cite{Mahanta4128}. For 
the radion mass $m_\phi \lesssim 100$ GeV, the bounds from the SM Higgs search 
\cite{Bae} are helpful, since $\Lambda_{\phi} \lesssim 600$ GeV can be
excluded. The limits from neutrino phenomenology and TASSO and CLEO on 
inclusive $B$ decays give lower bounds of 
the radion mass about 1 GeV \cite{Mahanta6006}, which agree to 
our lower bound. And the region $\Lambda_{\phi} \sim 750$ GeV for cutoff
$\Lambda=1$ TeV is excluded by the constraint from the weak mixing angle 
\cite{Kim}, since $e^{-kr_c\pi}=O(\Lambda_{\rm EW}/\Lambda_{\rm Pl})$
and $v \simeq 246$ GeV make $m_\phi$ less than about 260 GeV. 
Most of the previous constraints gives only 
lower bounds. Only the unitary bound in the $W^+_{L}W^-_{L} \rightarrow h \phi$
process gives the weak upper bound of $m_\phi$. Compared with the above
constraints, the $e^{-kr_c\pi}=O(\Lambda_{\rm EW}/\Lambda_{\rm Pl})$
and $v \simeq 246$ GeV constraints can produce the bounds of $m_\phi$ and
$\Lambda_\phi$ strongly, except the mass lower bound. 

%%%%%%%%%%%%%%%%%%%%%%%%
\section{Conclusions}
%%%%%%%%%%%%%%%%%%%%%%%%
Assuming the Goldberger-Wise mechanism in the Randall-Sundrum theory, we
considered the allowed regions of $m_{\phi}$ and $\Lambda_\phi$ from
two constraints, warp factor $e^{-kr_c\pi}=O(\Lambda_{\rm EW}
/\Lambda_{\rm Pl})$ and Higgs $vev$ $v \simeq 246$ GeV, by means of the 
tree level and one-loop level potentials. The
allowed regions are $0.4~ {\rm GeV} \lesssim m_{\phi} \lesssim 49~ {\rm GeV}$
and $760~ {\rm GeV}\lesssim \Lambda_{\phi} \lesssim 5900~ {\rm GeV}$ 
for the tree level potential, and  $0.8~ {\rm GeV} \lesssim m_{\phi} \lesssim
260~ {\rm GeV}$ and $1400~ {\rm GeV} \lesssim \Lambda_{\phi} 
\lesssim 1500~ {\rm GeV}$ for the one-loop level potential.
%with the region below the line 
%$\Lambda_{\phi} \approx 6m_{\phi}$ forbidden for the one-loop level potential. 
It is phenomenologically important that the one-loop allowed upper bound of 
radion mass $m_\phi$ is considerably larger than the tree-level one, but
the radion is still the first experimental 
signature of the RS model, since it is 
lighter than the KK modes. The mass of the radion is
less than about 260 GeV, and thus it decays into gluon or $W$ boson 
pairs dominantly. And its loop contributions are small 
due to the average value of the effective coupling constant $v/\Lambda_\phi 
\sim 1/6$ (the average effective fine structure constant is about $2 \times
10^{-3} \lesssim \alpha_{\rm em}$). The conditions of $e^{-kr_c\pi}
=O(\Lambda_{\rm EW}/\Lambda_{\rm Pl})$ and $v \simeq 246$ GeV can 
give the strong bounds of $m_{\phi}$ and $\Lambda_{\phi}$.
%than other constraints, except the mass lower bound. 
\\

{\bf Note Added:} After completing this paper, we received an interesting paper
by U. Mahanta \cite{11148}, who considered the one-loop contributions of 
the radion and the KK modes of graviton to the Higgs potential. We did not
include the radion contribution due to the non-renormalizability in Section II. 
In Ref. \cite{11148}, the one-loop contributions via some approximations are
too small to cause any instability of the classical vacuum in the valid region
of perturbation theory. Therefore, these contributions do not
change our results significantly. 

%%%%%%%%%%%%%%%%%%%%%%%%%%%
\section*{acknowledgement}
%%%%%%%%%%%%%%%%%%%%%%%%%%%
We would like to thank Prof.~Pyungwon Ko for discussions.
This work is supported by the Brain Korea 21 Project and grant No. 
1999-2-111-002-5 from the interdisciplinary research program of the KOSEF.

%%%%%%%%%%%%%%%%%%%%%%%%%%%%%%%%%%%%%%%%%%
\newpage
%%%%%%%%%%%%%%%%%%%%

%%%%%%%%%%%%%%%%%%%%%%
%Figure
%%%%%%%%%%%%%%%%%%%%%%
%%%Fig. 1

%%%Fig. 2

%%%Fig. 3

%%%%%%%%%%%%%%%%%%%%%%%%%%%
%%%%%%%%%%%%%%%%%%%%%%%%%%%
\end{document}